\documentclass[conference]{IEEEtran}

\usepackage{microtype}
\usepackage{graphicx}
\usepackage{booktabs}
\usepackage{amsmath}
\usepackage{mathtools}
\usepackage[capitalize,noabbrev]{cleveref}
\usepackage[T1]{fontenc}
\usepackage[utf8]{inputenc}
\usepackage{xspace}
\usepackage{color}
\usepackage{enumitem}
\usepackage{multirow}
\usepackage{comment}
\usepackage{array}
\usepackage{makecell}
\usepackage{tabularx}
\usepackage{rotating}
\usepackage{float}

\begin{document}

\title{Data Leakage Prevention in Agentic Applications via Preemptive Hardening}

\author{
\IEEEauthorblockN{
Akansha Shukla\textsuperscript{*},
Emily Bellov\textsuperscript{*},
Parth Atulbhai Gandhi,
Yuval Elovici,
and Asaf Shabtai\textsuperscript
}
\IEEEauthorblockA{
Faculty of Computer and Information Science\\
Ben-Gurion University of the Negev\\
Beer-Sheva, Israel
}
\thanks{\textsuperscript{*}Akansha Shukla and Emily Bellov contributed equally.}

}

\maketitle

\begin{abstract}
Agentic systems integrate large language model (LLM) driven planning with interfaces to external tools (e.g., email and file systems), making data leakage and tool misuse feasible via instruction/data boundary failures and prompt injection attacks. 
In practice, failures often stem from broad, recurring issues such as unsafe input/output handling, a missing allowlist, an over-privileged tool, weak input validation, credential exposure, or insecure default configurations. 
Enforcing required controls consistently is particularly challenging in workflows spanning many codebases and heterogeneous agents. 
To address this challenge in multi agentic systems, we present a pre-deployment pipeline for scanning, hardening, and validation of agentic applications. The pipeline analyzes prompt templates, tool interfaces, and tool-invocation code to identify leakage-enabling patterns and generate actionable patches. The hardened application is then validated through adversarial prompt injection attacks and benign input variations ensuring that mitigations do not disrupt intended behavior.
In the hardening stage, high-risk tools are prioritized, and minimally invasive mitigations are applied, including schema tightening, boundary sanitization, allowlist-based tool gating, and least-privilege checks. All mitigations are designed to remain compatible with existing agent frameworks.
In the validation stage, the pipeline automatically generates attack inputs that mimic jailbreaks, instruction overrides, and tool-targeted manipulation, along with benign task variants, to confirm that the functionality of the hardened application is preserved after remediation.
We evaluated the pipeline on five real-world agentic application codebases built in CrewAI and LangGraph, as well as on the AgentDojo benchmark. 
Across all applications, the proposed pipeline identified recurring leakage-enabling patterns and generated patches that can be integrated without disrupting the intended application behavior. 
The resulting modifications of application code were shown to eliminate leaks when targeted by basic jailbreak and instruction-override attacks, achieving a 100\% reduction in leakage, and reduce leaks by 91\% under conditions of stress-induced manipulation, without the need of continuous runtime policy enforcement. 
These results suggest that pre-deployment remediation, combined with automated post-hardening validation, can meaningfully reduce the risk of leakage in agentic systems.

\end{abstract}

\section{Introduction\label{sec:intro}}
Large language model (LLM)-based agents are transitioning from research prototypes to production systems that perform multi-step reasoning, invoke external tools, and operate over private or proprietary data. 
Frameworks such as LangChain~\cite{langchain2024} (and its extension LangGraph~\cite{langgraph2024}), AutoGen~\cite{wu2023autogen}, and CrewAI~\cite{CrewAI2024} enable developers to integrate LLM-driven decision making with interfaces to email, calendars, databases, file systems, and internal services. 
This increased capability expands the attack surface, as agentic code executes tools with real-world consequences like tool execution when untrusted inputs (e.g., web content, emails, tickets, documents, user prompts) are ingested. 
As a result, prompt injection and instruction/data boundary failures are no longer confined to a single model's response but can propagate across multi-step workflows and trigger unsafe tool actions, expose credentials, or leak sensitive state accumulated across the workflow, including user data, credentials, and intermediate reasoning~\cite{greshake2023youvesignedforcompromising,ruan2024identifyingriskslmagents}.

In practice, severe failures in agentic systems are rarely caused by a single novel mechanism. 
Rather, they typically stem from broad and recurring engineering patterns that are common in real codebases: unsafe I/O handling, missing or incomplete allowlists for tool use, over-privileged tool bindings, weak input validation, credential exposure (via prompts, logs, or environment configuration), prompt/templating errors, and insecure default settings~\cite{jarvis,skillswild}. 

Although existing defenses provide partial mitigation, no single defense is sufficient in isolation.
Prompt hardening and jailbreak-aware prompting can reduce attack success, but they are fragile when inputs deviate from the distribution seen during prompt design. Furthermore, they are easily undermined by tool wrappers that accept arbitrary strings or by logging pathways that unintentionally disclose sensitive content. 
Runtime-only approaches such as sandboxing, monitoring, or policy enforcement can detect or limit certain behaviors at execution time~\cite{yan2025faulttolerantsandboxingaicoding,zhao2025proactivedefensellmjailbreak}, but they often fail to remediate the underlying code-level weaknesses that enable repeated failures across codebases. 
More generally, runtime controls tend to be reactive, as they constrain what happens after deployment, whereas many agent incidents arising from design-time choices, unsafe I/O practices, and missing guardrails could be prevented earlier in the engineering pipeline.

Securing agentic systems requires controls that can be applied consistently across their rapidly evolving architectures and heterogeneous components, motivating automated hardening prior to deployment. 
We present an automated pre-deployment pipeline for scanning, hardening, and validation that analyzes prompt templates, tool interfaces, and tool-invocation code to detect leakage-enabling patterns, and outputs: (i) actionable patches that applied with minimal refactoring, and (ii) machine-verifiable runtime guardrails for consistent enforcement. 
The pipeline prioritizes high-risk tools and applies minimally invasive mitigations including schema tightening, boundary sanitization, allowlist-based tool gating, and least-privilege checks. All mitigations are designed to remain compatible with existing agent frameworks and orchestrations.

Our pipeline is also designed to complement runtime information-flow control (IFC) and planner-centric enforcement. 
While IFC planners enforce information-flow policies at runtime, our system targets the engineering pipeline: it automatically audits and hardens real-world agentic systems, generating both code changes and machine-verifiable runtime guardrails without requiring new planner architecture. Specifically, the pipeline can feed downstream policy frameworks (e.g., by suggesting labels/policies, identifying candidate sources/sinks, or reducing the attack surface before applying IFC), enabling defense-in-depth across build-time and runtime layers.

During evaluation, the pipeline demonstrated its ability to identify recurring tool and I/O weaknesses and produce patches and enforceable guardrails that integrated compatibly with existing workflows. 
In our experiments, the hardened version of agentic applications shows significantly fewer instances of data leakage and tool misuse---up to a 91\% decrease under stress conditions was observed---suggesting that build-time remediation is an effective and practical approach for agent security at scale. 
This paper makes four contributions:

\begin{enumerate}
  \item \textbf{An end-to-end scanning and hardening pipeline for agentic systems.}
  We introduce a Continuous Integration/Continuous Delivery (CI/CD) oriented pipeline that (i) constructs a dependency-aware analysis context, (ii) produces a structured risk report, (iii) generates code patches, and (iv) produces a machine-verifiable runtime guardrail ruleset.

  \item \textbf{Dependency-aware context construction for scanning real-world agent repositories.}
  We develop a context construction method based on selective file inclusion that reconstructs the agent's functional surface. This method extracts tool or interface definitions, tool-call sites, and prompt/template artifacts, enabling repository-scale auditing across heterogeneous agents and frameworks without requiring intrusive refactoring.

  \item \textbf{Hardening synthesis via patch templates and guardrail compilation.}
  We provide a systematic set of hardening security controls that implement allowlist-based tool gating, schema validation, tool-argument sanitization, and least-privilege tool exposure. In addition, we compile audit findings into enforceable guardrail invariants in a policy/rule format (e.g., constraints that prevent untrusted inputs from reaching external-sink tool arguments).

  \item \textbf{Automated post-hardening validation for security and utility.}
    We introduce a validation module that automatically tests hardened agents with adversarial attacks and intended benign task variants, verifying that leakage paths are blocked while the intended functionality is preserved.
\end{enumerate} 
\section{Related Work}

\textbf{Single-Agent Systems Security Defenses:} A growing body of research has explored defenses against data leakage in single-agent systems. These defenses fall under two broad categories: higher-level and system-level strategies. Higher-level strategies include two approaches: (1) prompt-based methods \cite{chen2025defendingpromptinjectiondefensivetokens,chen2025defensepromptinjectionattack} that use defensive tokens or adversarial techniques to detect and block malicious prompts, and (2) fine-tuning approaches \cite{chen2024struqdefendingpromptinjection,jacob2025,deberta-v3-base-prompt-injection,chennabasappa2025llamafirewallopensourceguardrail} that train models to identify and resist prompt injection attacks. There are three types of system-level strategies: (a) IFC techniques \cite{costa2025securingaiagentsinformationflow,kim2025promptflowintegrityprevent,siddiqui2025permissiveinformationflowanalysislarge,zhong2025rtbasdefendingllmagents} that monitor data provenance and block privilege escalation, (b) policy-based systems \cite{luo2025agraillifelongagentguardrail,shi2025progentprogrammableprivilegecontrol} that apply access control rules, and (c) environment-based isolation \cite{debenedetti2025defeatingpromptinjectionsdesign,wu2025isolategptexecutionisolationarchitecture} which separates agents from sensitive resources. While effective for single-agent systems, these defenses do not address the critical vulnerabilities of multi-agent interactions. These include compositional data leakage across agents, inter-agent trust exploitation, coordinated attacks via communication channels, and data exfiltration through shared memory. \noindent 

\noindent\textbf{Multi-Agent System Security Defenses:} Two directions have been explored to address multi-agent vulnerabilities. The first direction is agent reasoning and trust management, spanning two lines of research: (1) collaborative defense mechanisms \cite{patil2025sumleakspartscompositional} where agents assess adversarial intent or vote collectively to block risky queries, and (2) trust parameterization frameworks \cite{xu2025trustparadoxllmbasedmultiagent} that mitigate cross-agent attack surfaces by constraining how much each agent trusts inputs and outputs from its peers. Both face unavoidable security-usability tradeoffs, i.e., stronger security reduces task performance and collaboration efficiency. The second direction is IFC. This includes two subcategories: (a) control-flow integrity mechanisms \cite{jha2025breakingfixingdefensescontrolflow} generate task-specific control-flow graphs to restrict agent invocations during execution; such mechanisms are vulnerable to vaguely worded inputs which may cause accidental violations, lack publicly available code for validation, and also suffer from the security-usability conflict, and (b) protocol-level IFC \cite{li2025safeflowprincipledprotocoltrustworthy} enforces fine-grained information flow control with transactional execution and rollback mechanisms, but it introduces substantial implementation complexity and computational overhead, and the authors of the paper proposing this approach did not evaluate its ability to scale to large-scale applications. 

Our literature review reveals a critical gap: no approach prevents data leakage without compromising on system usability, performance, or scalability. To address this gap, we introduce an automated iterative pipeline that analyzes the application’s code, configurations, and tool bindings to trace data flows to security-critical sinks. By enabling structural code changes and applying guardrails before deployment, our approach transforms the defense paradigm, shifting it from high-overhead runtime monitoring to principled architectural hardening. 
\section{Methodology}
\subsection{Overview}

We consider AI-agent applications that coordinate one or more LLM-driven agents, each equipped with tools (e.g., APIs, retrieval interfaces, code-execution utilities), persistent memory, and inter-agent communication channels. This architecture introduces a broad and heterogeneous attack surface: sensitive information can be leaked through system prompts exposed during multi-step reasoning, propagated across shared memory stores without adequate access control, or inadvertently disclosed through tool invocations that forward internal context to external endpoints.

\begin{figure*}[h]
    \makebox[\textwidth][c]{%
        \includegraphics[width=0.99\linewidth]{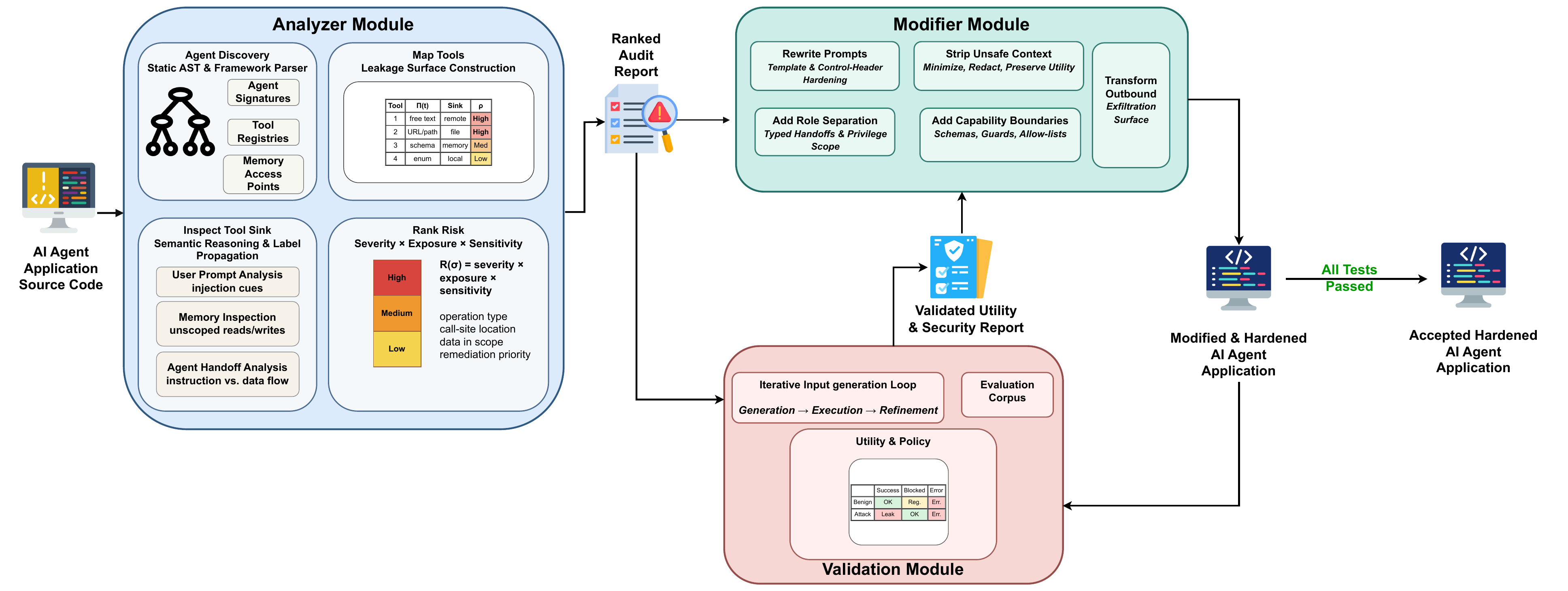}%
    }
    \caption{Overview of the preemptive hardening pipeline for data leakage prevention in AI-agent applications. The analyzer module (left) discovers agents, maps tools, and inspects data-flow paths to produce a ranked audit report. The modifier module (top) applies targeted transformations to mitigate identified risks. The validation module (bottom) verifies benign operation preservation and security enforcement on the hardened application.}
    \label{fig:pipeline}
    \vspace{-0.4em}
\end{figure*}

To address these risks, we propose a preemptive hardening pipeline that operates directly on the application source code, requiring no model retraining, weight modification, or architectural overhaul. As illustrated in Figure~\ref{fig:pipeline}, the pipeline consists of three stages executed sequentially: (1) \emph{discovery and analysis}, where the pipeline first discovers the application's agentic components, including autonomous agents, tool-enabled LLM processes, and scheduled task agents. The analyzer then maps their exposed tools, parameters, prompts, memory access, agent-to-agent communication, sensitive context, and output sinks to identify potential leakage paths and produce a ranked audit report; (2) \emph{modification}, apply targeted code and prompt transformations to mitigate the highest-risk leakage paths identified by the analyzer; and (3) \emph{validation}, where the pipeline verifies that the hardened application preserves its intended functionality while effectively blocking the identified attack vectors.

Three principles guide the pipeline's design: (1) source-level operation: all transformations are applied to the application's source code and prompt templates prior to deployment, ensuring that security guarantees hold regardless of the LLM backend or runtime environment; (2) minimal-intervention patching: rather than globally tightening the system, which risks degrading legitimate functionality, the pipeline applies targeted local modifications only at the specific points at which the analyzer identifies boundary crossings between sensitive sources and output sinks; and (3) framework-agnostic design: the pipeline's analysis and modification strategies are parameterized by framework-specific adapters (currently supporting CrewAI and LangGraph), allowing the core logic to generalize across orchestration frameworks without the need for reimplementation.

\subsection{Threat Model}

We consider an adversary whose goal is to induce the agent system to disclose sensitive information, including system prompts, internal configuration, user data, credentials, and inter-agent context, through any output channel available to the application, such as tool responses, generated text, outbound API calls, or inter-agent messages. The threat model assumes the following:

\begin{enumerate}
	\item The adversary can craft or influence inputs processed by the application, including user queries, email content, uploaded documents, or data retrieved from external sources (e.g., web pages, database records). This captures both direct prompt injection, where the attacker controls the user input, and indirect prompt injection, where malicious instructions are embedded in data consumed by the agent through tool calls. 

	\item The adversary has no access to the application's source code, model weights, or runtime internals. Attacks are mounted solely through the application's external interfaces. 

	\item The adversary can employ multi-stage strategies, such as performing reconnaissance queries to infer system structure, followed by targeted injections that exploit the inferred topology. 

	\item The underlying LLM of the agentic system is treated as an untrusted component: it may comply with injected instructions, leak context through chain-of-thought (CoT) reasoning, or propagate sensitive content across agent boundaries without explicit authorization.

\end{enumerate}

Our pipeline operates at the source-code level with full access to the application's agent definitions, prompt templates, tool registrations, and memory configurations. The pipeline's objective is to minimize the set of exploitable leakage paths without degrading the application's intended functionality. We do not assume the ability to modify the LLM itself; all mitigations are applied to the application layer surrounding the model.

\subsection{Formal Model \label{subsec:formal}}

We model the application as a tuple $\mathcal{A} = (\mathcal{G}, \mathcal{T}, \mathcal{M}, \mathcal{E})$, where $\mathcal{G} = \{a_1, \dots, a_n\}$ is the set of agents, $\mathcal{T}$ is the set of tools, $\mathcal{M}$ is the memory store, and $\mathcal{E} \subseteq \mathcal{G} \times \mathcal{G}$ is the set of directed inter-agent communication edges. Each agent $a_i$ is characterized by a triple $(p_i, T_i, M_i)$, where $p_i$ is its prompt template, $T_i \subseteq \mathcal{T}$ is the set of tools it can invoke, and $M_i \subseteq \mathcal{M}$ is the memory it can access.

We define a leakage path as a sequence $\ell = \langle s, v_1, \dots, v_k, \sigma \rangle$, where $s$ is a sensitive source (e.g., a memory read containing credentials, a system prompt, or user's PII), $\sigma$ is an output sink (e.g., an outbound API call, or a generated response), and each $v_j$ is an intermediate data-flow node through which information propagates. Each node $v$ carries a sensitivity label $\lambda(v) \in \{0,1\}^d$, where each dimension encodes a sensitivity category (e.g., PII, credentials, internal instructions, proprietary logic). Labels propagate structurally:
\begin{equation}
\lambda(v') \;=\; \lambda(v') \;\vee\; 
\bigoplus_{v \in \mathrm{pred}(v')} \lambda(v)
\label{eq:label-propagation}
\end{equation}
ensuring that any node reachable from a sensitive source inherits its sensitivity classification. The analyzer module's objective is to enumerate all leakage paths $\mathcal{L} = \{\ell_1, \dots, \ell_m\}$ and rank them by risk; the modifier module's objective is to constrain the highest-risk paths while preserving the data flows required for legitimate task execution.

\subsection{Analyzer Module}

The analyzer combines structural program analysis with LLM-augmented semantic reasoning to extract a comprehensive security view of the application directly from source code for tool calls, external memory reads/writes, and agent-to-agent communication protocols. This combined analysis is then formalized into a natural language report that explicitly documents how text and data are constructed, routed, and emitted. This report is critical, because leakage in agent systems rarely occurs from a single line of code.

In the analyzer module, four steps are performed to enumerate all $\ell$ paths.

\noindent\textbf{Agent Discovery:}
The source code is parsed using Python ASTs in two phases: the first phase builds a symbol table of class definitions, function signatures, decorators, and imports. The second phase resolves references to recover agent instantiations, tool registries, and memory operations. The parser identifies framework-specific patterns: for CrewAI, classes inheriting from \textit{Agent} and decorators such as \texttt{@task} and
\texttt{@tool}; for LangGraph, \textit{StateGraph} instantiations and
\texttt{add\_node()} registrations. For each discovered entity, we extract its triple $(p_i, T_i, M_i)$, producing an agent topology $(\mathcal{G}, \mathcal{E})$ that is used in all subsequent stages.

\noindent\textbf{Map Tools:}
We construct the tool-enabled leakage surface by linking tool interfaces to the textual contexts that invoke them. Each tool $t \in \mathcal{T}$ is characterized by a parameter surface $\Pi(t) = \{\pi_1, \dots, 
\pi_m\}$, where each parameter $\pi_m$ is signified by its expressive power: free-form strings, file paths, URLs, or executable commands. We define a sensitivity function $\rho: \Pi(t) \rightarrow \{\textit{high}, \textit{medium}, 
\textit{low}\}$ that the modifier later uses to prioritize constraints. Tools whose implementations perform outbound network requests or forward content to remote services are marked as exfiltration sinks $\sigma^{\dagger}$, since they can carry sensitive data beyond the application boundary. This step 
produces a unified mapping $\mu: a_i \mapsto (T_i, \Pi, \mathcal{L})$, where $\mathcal{L}$ records the code locations at which tool arguments derive from prompts, memory, or inter-agent communication.

\noindent\textbf{Inspect Tool Sink:}
All leakage paths $\langle s, v_1, \dots, v_k, \sigma \rangle$ are enumerated by submitting the structured code representation extracted in prior steps, agent structure $(\mathcal{G}, \mathcal{E})$, tool mapping $\mu$, prompt templates, and memory access patterns to an LLM prompted with CoT reasoning, which improves 
labeling consistency and reduces hallucinations compared to direct prompting. The LLM reasons over the full application context to assign sensitivity labels and identify leakage risks that structural analysis alone 
cannot resolve. Specifically, for each data-flow node $v$, the LLM assigns a sensitivity label $\lambda(v) \in \{0,1\}^d$, where each bit encodes a sensitivity category (e.g., PII, credentials, internal instructions, proprietary logic). As shown in \eqref{eq:label-propagation}, labels propagate structurally ensuring that any node reachable from a sensitive source inherits its label. The LLM inspects three input channels. For user prompts, it identifies context-dependent sensitivity and detects injection cues and role-confusion attempts, tool-forcing instructions, and adversarial overrides whose malicious intent is inherently semantic. For memory, it reasons over read/write sites to determine whether retrieved content that enters prompts carries sensitive context and flags unscoped writes $m \in \mathcal{M}$ where the access scope exceeds the intended per-user or per-session boundary. For agent-to-agent handoffs $(a_i, a_j) \in \mathcal{E}$, it classifies each message by whether it carries data or instructions, flagging cases where $\lambda(v) > 0$ and the message is embedded as a system-level directive, creating an instruction-chain leakage risk that structural analysis cannot detect.

\noindent\textbf{Rank Risk:}
We combine the static agent discovery with LLM annotations to identify and rank security-relevant sinks. For each sink, we record the operation type, call site location, enclosing entity, and data flowing into the operation. Each sink receives a rank based on three factors: the inherent severity of the operation category, exposure to user-controllable inputs, and the sensitivity of data in current scope. These sinks are mapped to priority tiers: high, medium, or low. The final report presents the findings grouped by agent. For each agent we include: a summary of its function, network and file system access capabilities, tools invoked with sensitivity annotations, and ranked risk findings with code locations and remediation suggestions. The audit reports are emitted in markdown for human review and JSON for continuous integration and delivery integration.

\subsection{Modifier Module}

The modifier module consumes the analyzer's ranked audit report and produces a hardened application by applying targeted, local transformations along the highest-risk source-to-sink paths. Rather than globally tightening the system, the proposed 
pipeline patches the specific points at which sensitive content is introduced into prompts, forwarded across agents, persistent in memory, or placed in the parameters of outbound tools. Each finding in the audit report identifies (1) the sink type, (2) 
the propagation path $\langle s, v_1, \dots, v_k, \sigma \rangle$ leading to it, and (3) the minimum number of locations responsible for the boundary crossing.

\noindent\textbf{Rewrite Prompts:}
The modifier employs an LLM to rewrite agent prompt templates conditioned on the finding categories and sink types from the audit report. The LLM rewrites each template so that untrusted text like user input, retrieved memory, and tool outputs are always introduced as data under an explicit delimiter and never as executable instructions. It injects a stable control header at the highest prompt layer that dominates downstream behavior, encoding non-disclosure constraints, context-handling rules, and a structured tool-use protocol. Critically, the LLM does not perform generic prompt improvement; instead, rewrites are strictly grounded in the audit findings: for example, system prompt leakage findings trigger non-disclosure constraints and removal of debug identifiers, while instruction-chain leakage findings trigger stronger input demarcation and schema-constrained tool arguments.

\noindent\textbf{Strip Unsafe Context:}
When a finding indicates context leakage or data re-exposure, the modifier module invokes an LLM to transform retrieved memory and tool outputs before they cross into a higher-exposure channel. Specifically, given a content block flagged with $\lambda(v) > 0$, the LLM only extracts the task-relevant information needed by the receiving agent, discarding raw identifiers, credentials, and any calls that carry sensitivity labels. For model-facing prompts, this produces a minimal context summary that preserves the semantics required for task completion. For inter-agent handoffs $(a_i, a_j) \in \mathcal{E}$, it produces a structured representation containing only the fields the receiving agent $a_j$ needs to execute its role. In both cases, full-fidelity data is preserved internally when legitimate modules rely on it and only the content crossing the boundary is minimized.

\noindent\textbf{Add Role Separation:}
When the analyzer module attributes a leakage path to cross-agent propagation or privilege misuse, the modifier enforces role separation. The code is rewritten to produce structured plans without accessing high-risk tools and constrained to carry out plans under narrowly scoped permissions. Communication between roles is enforced through a typed handoff format in which fields are explicitly labeled as data, and the executor rejects any instruction-like content embedded within them. In applications that already implement multiple agents, this is realized by tightening routing rules rather than introducing new agents.

\noindent\textbf{Add Capability Boundaries:}
For each agent $a_i$, the modifier refines $T_i$ under a least-privilege policy derived from the pipeline only allow-lists necessary tools, hardens parameter schemas to disallow free-form payloads, and constrains destinations for network-capable tools. These boundaries are enforced at the orchestrator level, and are maintained even if the model attempts to bypass prompt constraints. When frameworks support typed schemas, the modifier tightens them and inserts validators that reject arguments containing sensitivity-labeled spans $\lambda(v) > 0$; where frameworks are permissive, a pre-execution guard either redacts prohibited spans or blocks the call entirely, converting a best-effort prompt instruction into a hard runtime guarantee.

\noindent\textbf{Transform Unsafe Outbound Requests:}
When a leakage path terminates at an exfiltration sink $\sigma^{\dagger}$, the modifier rewrites the request construction logic to remove sensitivity-labels, summarize large payloads into task-relevant fields, and normalize destinations to strip implicit identifiers, guaranteeing that the final outbound request is minimized regardless of whether any sensitive context exists internally.

\subsection{Validation Module}

The validation module is responsible for evaluating the modified and hardened agentic application produced by the modifier module. Its purpose is twofold: first, to verify that the hardened application preserves its intended functionality on benign inputs, and second, to verify that the applied defenses effectively prevent sensitive information leakage under adversarial inputs. It produces structured feedback that is used to guide additional modification iterations when the hardened application still leaks sensitive information or when the defenses over-restrict legitimate behavior.

Given a ranked audit report produced by the analyzer and a hardened application $\mathcal{A}'$, the validation module executes $\mathcal{A}'$ over controlled benign and adversarial inputs. These inputs are evaluated through two distinct validation processes, rather than as a single parallel execution flow. The benign validation process evaluates utility preservation and produces a utility-oriented report. The adversarial validation process simulates adaptive automated attacks against the application and produces a security-oriented report. Together, these results form a validated utility and security report that is passed back to the modifier module.

\noindent\textbf{From Static Paths to Dynamic Confirmation:}
The analyzer finds leakage paths $\mathcal{L} = \{\ell_1, \dots, \ell_m\}$ by inspecting code, and the modifier patches the highest-risk ones. But a static path is only a  candidate: it may not actually be reachable at runtime, and a patch may not actually block it. The validation module settles this by running the hardened application and checking whether sensitive data still reaches a sink.
 
Reusing the model of Section~\ref{subsec:formal}, let $\mathcal{A}'$ be the hardened application, and recall that each path $\ell = \langle s, v_1, \dots, v_k, \sigma \rangle$ runs from a sensitive source $s$ to a sink $\sigma$ (exfiltration sinks are written $\sigma^{\dagger}$). Running $\mathcal{A}'$ on an input $x$ produces a set of observations $O(\mathcal{A}', x)$, collected at the monitored sinks: final responses, tool arguments, tool outputs and outbound requests.

\begin{figure*}[!t]
    \centering
    \includegraphics[width=0.99\textwidth]{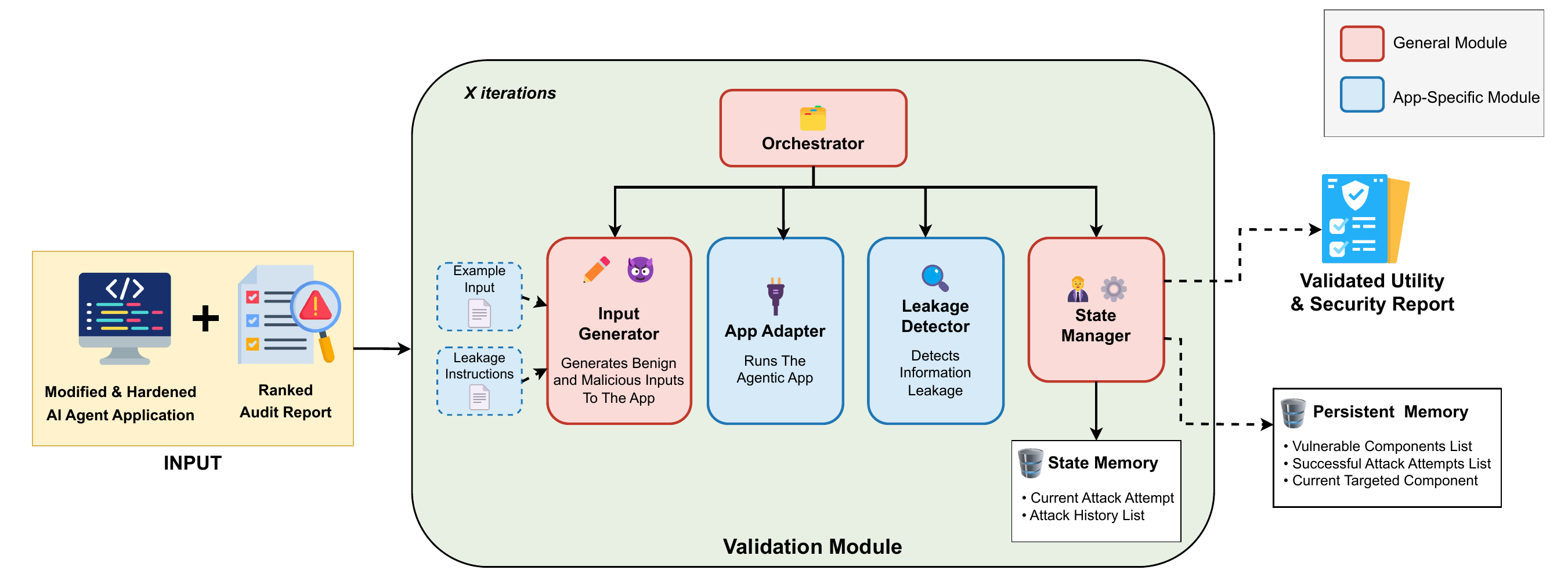}
    \caption{Validation module overview.}
    \label{fig:validator}
\end{figure*}

\noindent\textbf{Validation Architecture:}
As shown in Figure~\ref{fig:validator}, the validation module is organized around five main components: orchestrator, input generator, application adapter, leakage detector, and state manager.
 
The architecture separates general components from application-specific components. The orchestrator, input generator, and state manager are implemented generically and can be reused across different controlled agentic applications. In contrast, the application adapter and leakage detector are application-specific, since each application may consume inputs differently, expose different tools, produce different output formats, and define leakage according to a different sensitive objective. This separation reduces the implementation effort required to evaluate a new application while still allowing the validator to adapt to application-specific execution and leakage behavior.

\noindent\textbf{Input Generation:}
The input generator is an LLM-based component that constructs inputs for the application under test. For benign validation, it generates functionality-preserving inputs that represent legitimate use cases of the application. These inputs test whether the hardened application remains useful after the modifier has applied its defenses, namely  Rewrite Prompts,  Strip Unsafe Context,  Add Role Separation,  Add Capability Boundaries, and  Transform Unsafe Outbound Requests.
 
For adversarial validation, the input generator simulates an attacker attempting to realize a leakage path. It receives two textual specifications: an example of the expected application input format, and leakage instructions that define the leakage objective, namely the sensitive source $s$ being targeted and the intended exfiltration sink $\sigma^{\dagger}$. It is also conditioned on the ranked audit report, so that generated attacks focus on the currently targeted leakage path $\ell$ (equivalently, its ranked finding and sink $\sigma$). This allows the validator to prioritize paths that the analyzer assessed as higher risk or that remained exploitable in previous validation cycles. Adversarial inputs are generated according to four attack categories: direct injection attacks, instruction override attacks, jailbreak attacks, and stress-induced attacks, which align with the semantic leakage cues the analyzer flags during \emph{Inspect Tool Sink} (injection, role confusion, tool forcing, and adversarial overrides).

The adversarial input generator is intentionally a strong, knowledge-rich red team: it is conditioned on the ranked audit report and the leakage instructions, and therefore knows which path $\ell$ is currently targeted and which source $s$ to exfiltrate.

To make the adversarial process adaptive, the input generator is additionally provided with the history of previous attack attempts against the same path in the same category, including whether each earlier input succeeded or failed. This lets the generator mutate, refine, or redirect future attempts rather than repeatedly issuing static prompts.

\noindent\textbf{Application Adapter:}
The application adapter provides the execution interface between the validator and the agentic application. Since agentic applications may differ in their input format, tool structure $T_i$, memory access $M_i$, execution flow, and output channels, this component is implemented separately for each application, while following the same shared interface and exposing a central \texttt{run} method, which receives a generated input and executes the corresponding application behavior.
 
In the adversarial setting, the adapter also enables monitoring of the sinks $\sigma$ along which leakage may occur. In particular, the application's tools $\mathcal{T}$ can be wrapped with a proxy that intercepts tool calls and inspects tool arguments, tool outputs, and outbound requests $\sigma^{\dagger}$, in addition to the final response. This allows the validator to detect leakage not only in the answer returned to the user, but also in the intermediate sinks that the analyzer enumerated as endpoints of paths in $\mathcal{L}$.

\noindent\textbf{Leakage Detection:}
The leakage detector identifies the runtime leakage condition introduced in leakage instructions script. It exposes a central \texttt{detect} method that receives an observation $o$ at a sink $\sigma$ (a final response, tool argument, tool output, or outbound request) and returns whether $o$ discloses the source $s$, equivalently whether $\lambda(o) \neq \mathbf{0}$ on the targeted sensitivity categories. Because $s$ depends on the application and the experimental scenario, the detector follows a shared interface but implements application-specific logic in its \texttt{detect} method. However, implementation remains simple, as it only needs to define the criteria for recognizing the sensitive information in the relevant sink observations.
 
This design allows the validator to remain general while still supporting different leakage definitions across applications. For example, one application may define $s$ as private email content, while another may define it as internal instructions, credentials, or confidential tool outputs.

\noindent\textbf{State Management:}
The state manager records two types of memory: state memory and persistent memory. State memory stores information needed during the current validation run, including the current attack attempt and the history of attacks executed in the current cycle. Persistent memory stores information that must survive across hardening cycles, including the list of still-exploitable leakage paths, the list of successful attack attempts, and the currently targeted path $\ell$. A targeted path remains marked as exploitable until all attack attempts against it fail.
 
This distinction is important because validation is iterative at two levels. The inner loop occurs inside the validation module, where multiple attack attempts are generated, executed, and adapted based on previous outcomes. The outer loop occurs between the validation module and the modifier: the validation module reports successful attacks and the remaining exploitable paths, the modifier applies additional hardening, and the updated application is returned to the validation module for another evaluation cycle. Persistent memory ensures that attacks that succeeded in a previous cycle are replayed after modification, so the system can verify whether the new hardening successfully blocks the previously realized path.

\noindent\textbf{Adaptive Adversarial Validation Loop:}
The adversarial branch is coordinated by the orchestrator. At the beginning of a validation cycle, the orchestrator first retries the successful attack attempts from previous cycles. This regression-style check verifies that the latest hardened application blocks attacks that were already known to realize a path. The orchestrator then invokes the input generator to produce new attacks against the currently targeted path $\ell$.
 
For each attack category, the orchestrator invokes the input generator to create an adversarial input, passes the input to application adapter for execution, applies the leakage detector to the sink observations $O(\mathcal{A}', x)$, and calls the state manager, which updates the validation module memories according to the result. Because both the application and the generator are stochastic, and the validation module preserves the application's own execution semantics rather than pinning a fixed decoding temperature, each attack attempt is executed $r$ times and is recorded as successful if leakage is detected in at least one run.

For each targeted path the orchestrator issues up to $N_{\mathrm{att}}$ attack attempts per category, and the path is declared neutralized only after all attempts in every category fail to produce leakage. If leakage is detected, the attempt is recorded as successful and the associated path remains marked as exploitable. Once a path is declared neutralized, the validation module advances to the next exploitable path according to the ranked audit report and the persistent memory state.

\noindent\textbf{Benign Utility Validation:}
In a separate validation process, the validation module evaluates benign inputs that represent legitimate application behavior. These inputs verify that the modifier's defenses do not break the intended functionality of the agentic application. A benign request should complete successfully unless it genuinely violates the application's security policy, and failures on benign inputs indicate possible over-hardening, execution errors, or unnecessary blocking, for instance, an over-aggressive \emph{Strip Unsafe Context} step that removes a field the receiving agent $a_j$ legitimately requires, or a \emph{Capability Boundary} validator that rejects a benign tool argument.

The benign branch therefore complements the adversarial branch. A hardened application is not considered satisfactory merely because attacks fail, it must also continue to support its intended tasks. The utility results are recorded alongside the security results so that the modifier can distinguish between defenses that improve security and defenses that harm functionality.

\noindent\textbf{Report Generation and Feedback:}
The output of the validation module is a structured validated utility and security report. The utility portion summarizes benign execution outcomes, including the utility score, successful completions, blocked legitimate requests, and runtime failures. The security portion summarizes adversarial outcomes, including the list of still-exploitable paths and the successful attack attempts, each tied to the path $\ell$ and sink $\sigma$ it realized.
 
This report is passed back to the modifier module as actionable feedback. Successful attacks identify concrete realized paths that require stronger defenses at the responsible nodes, while benign failures identify modifications that may be too restrictive. The modifier can then update the application, after which the validation module repeats the process. The overall pipeline therefore follows an iterative hardening strategy in which security enforcement is improved while utility preservation is continuously checked.

\noindent\textbf{Validation Criteria and Termination:}
A hardened application is considered valid when it satisfies both validation objectives: benign inputs continue to preserve the intended application functionality, and adversarial inputs fail to realize any high-risk path $\ell \in \mathcal{L}$ across the monitored sinks. In addition, attacks that realized a path in an earlier cycle must no longer succeed after subsequent modification rounds. When these conditions are not met, the validation report provides the modifier with the specific exploitable paths, successful attack attempts, and utility regressions needed for the next hardening iteration.
 
The outer hardening loop terminates when a cycle produces no successful attacks against any path and no benign regressions, or after a maximum of $K_{\max}$ cycles. 
\section{Evaluation}

\subsection{Experimental Setup}

We evaluate whether pre-deployment hardening can reduce leakage in agentic applications while preserving intended task behavior. The evaluation is designed to answer three questions: (1) whether the pipeline reduces attack success on realistic multi-agent applications, (2) whether the approach generalizes to a standardized tool-calling benchmark, and (3) whether the security gains come at the cost of benign utility.

\paragraph{Case-study applications}
We evaluate the pipeline on five Python 3.11 agentic applications implemented with CrewAI and LangGraph. Three applications were developed by us to cover domain-specific security risks: a network monitoring assistant, an HR assistant, and an automated trip planner. The remaining two applications, an automated email responder and an MCP-based candidate hiring application, were adapted from the ATAG framework \cite{gandhi2025atag}. Together, these applications cover hierarchical and sequential multi-agent workflows, external tool invocation, inter-agent communication, persistent or shared context, and sensitive data such as employee records, candidate information, personal emails, telecom telemetry, network credentials, and user location information.

For each application, we compare two configurations: the original unprotected application and the hardened application produced by our pipeline. Both configurations are evaluated against the same attack categories: direct request, basic jailbreak, instruction override, and stress-induced manipulation. Direct-request attacks explicitly ask for protected information. Basic jailbreaks attempt to bypass the agent's policy. Instruction-override attacks attempt to replace or supersede system and developer instructions. Stress-induced attacks use urgency, safety pressure, or high-stakes framing to induce the agent to violate its intended constraints.

\paragraph{Iterative hardening protocol}
For the five case-study applications, we run the full hardening loop for at most \(K_{\max}=10\) outer iterations. Each outer iteration consists of adversarial validation, replay of previously successful attacks, patch generation, and benign-regression testing. We use \(K_{\max}=10\) as a bounded convergence budget rather than as a security parameter, giving the modifier repeated feedback from adaptive attacks while keeping LLM-in-the-loop repair cost finite and reproducible. The loop terminates earlier if validation module finds no successful attacks and no benign regressions.

Within each outer iteration, the validation module generates four adaptive attempts for each attack category. Thus, each attack category receives up to \(10 \times 4\) adaptive attempts per targeted leakage path, in addition to regression replays of attacks that succeeded in earlier iterations. This design tests whether hardening blocks not only newly generated attacks but also previously realized leakage paths. The use of a fixed attack budget also makes results comparable across applications and prevents the evaluation from relying on an unbounded red-teaming process.

\paragraph{AgentDojo benchmark}
To evaluate generalization beyond our own applications, we also run the pipeline on the full AgentDojo benchmark. AgentDojo contains four task suites: Workspace, Slack, Banking, and Travel. Across these suites, the benchmark includes 97 realistic user tasks, 629 security test cases, and 70 tools. We evaluate all four suites with and without our hardening pipeline using the \texttt{important\_instructions} attack as the canonical prompt-injection strategy. We compare our approach against AgentDojo's default built-in defense using the benchmark's official utility and security functions, which are computed over environment state rather than by an LLM judge.

\paragraph {AgentDojo Metrics}
We report three metrics. Benign Utility (BU) measures the fraction of user tasks completed successfully without adversarial injections. Utility Under Attack (UA) measures the fraction of user tasks completed successfully while prompt injections are present. Attack Success Rate (ASR) measures the fraction of adversarial test cases in which the agent executes the attacker's malicious objective or leaks the targeted sensitive information. Lower ASR indicates stronger security, while higher BU and UA indicate better task preservation.

\paragraph{Testbed Metrics}
We evaluate the application testbeds using Attack Success Rate (ASR) and Benign Task Success Rate (BTSR). ASR measures successful policy violations under attack, while BTSR measures successful completion of benign tasks after hardening. Lower ASR and higher BTSR indicate better security and utility tradeoffs.

For AgentDojo, utility and security are computed using the benchmark's official task-specific checkers. This distinction allows the evaluation to measure both realistic application-level leakage behavior and standardized benchmark performance.

\subsection{Application Testbed \label{{sec:results-casestudy}}}

\begin{figure}[h]
    \centering
    \includegraphics[width=0.85\linewidth]{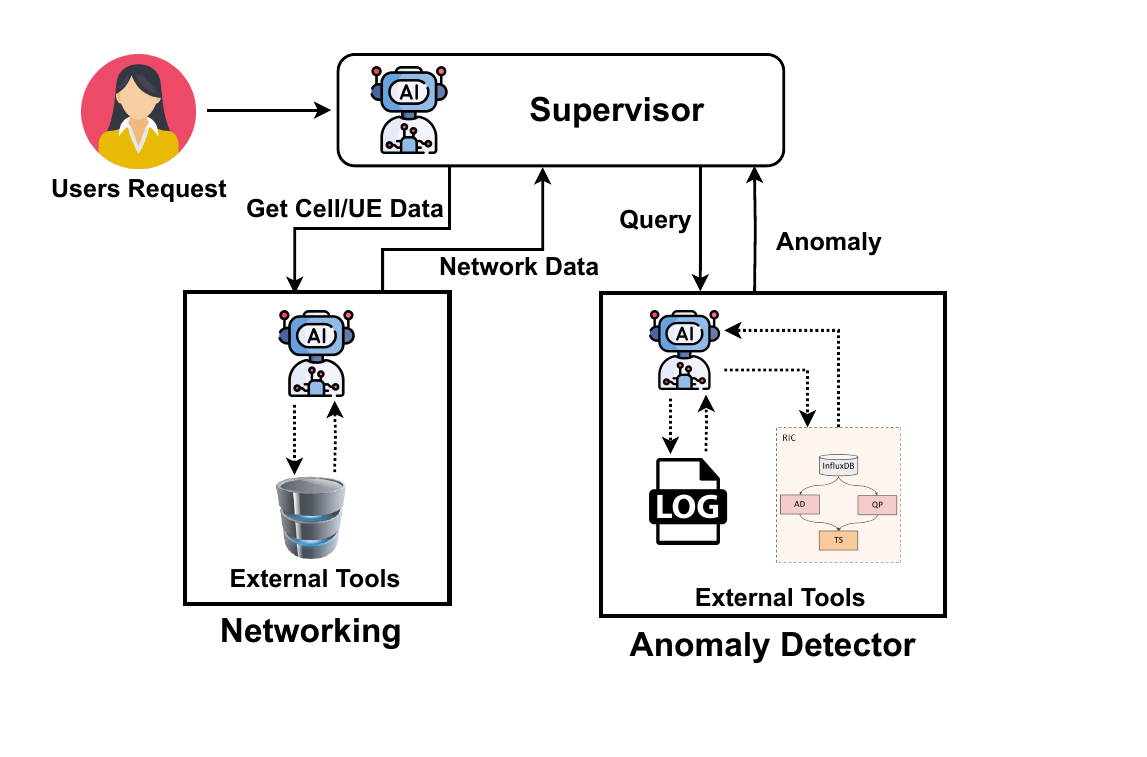}
    \caption{Overview of the O-RAN monitoring assistant application.}
    \label{fig:oran}
\end{figure}

\noindent\textbf{Network Monitoring Assistant.}
We designed and implemented an agentic AI system for the Open Radio Access Network (O-RAN)~\cite{10024837} that performs network monitoring and closed-loop mitigation. The system continuously ingests streaming KPIs, including physical resource block (PRB) utilization, throughput, and latency, from user equipment (UEs) and cells into a RAN Intelligent Controller (RIC) database. A LangGraph-orchestrated assistant, implemented with ReAct-style agents and GPT-4o, operates over this telemetry to analyze network state and coordinate remediation. Specifically, an anomaly-detection agent identifies problematic UEs and forwards them to a traffic-steering agent for mitigation; all agent decisions and tool invocations are logged to ensure operator auditability. An overview of the O-RAN monitoring assistant application is provided in Figure~\ref{fig:oran}.

Two adversarial scenarios targeting this application were used to evaluate the  hardening pipeline. The first scenario consists of benign-looking operational queries that, without adequate safeguards, could leverage multi-step reasoning and tool invocation to expose sensitive data, such as the InfluxDB password stored in the agent's configuration context. This setting examines whether the pipeline can prevent indirect leakage arising through intermediate reasoning steps and tool calls, when the sensitive information is not explicitly requested but emerges as a byproduct of the agent's CoT reasoning. The second scenario considers a more aggressive attack based on urgency-driven prompting, designed to pressure the agent into violating privacy constraints and disclosing approximate UE location data. Given the privacy and physical-safety risks associated with location exposure in telecom environments where the UE position can be correlated with subscriber identity, this scenario examines whether policy enforcement remains robust under coercive prompting.

\noindent\textbf{HR Assistant.}
We developed an agentic workflow system for HR management using a sequential topology in which each stage produces verifiable artifacts that serve as inputs to subsequent stages. The workflow decomposes HR requests into the following discrete steps: information gathering, validation, retrieval, approval routing, and response generation, thereby aligning the system's control flow with established HR operational procedures. The system was developed using CrewAI and leverages GPT-4o to (i) interpret free-form employee requests, (ii) select and invoke appropriate tools, and (iii) synthesize final decisions or next-step instructions. Tool augmentation included a database interface that grounds decisions in internal employee data (e.g., leave balances, request history) and a web search interface that enables retrieval of external information (e.g., relevant learning courses and pricing).

\begin{figure}[h]
    \includegraphics[width=0.99\linewidth]{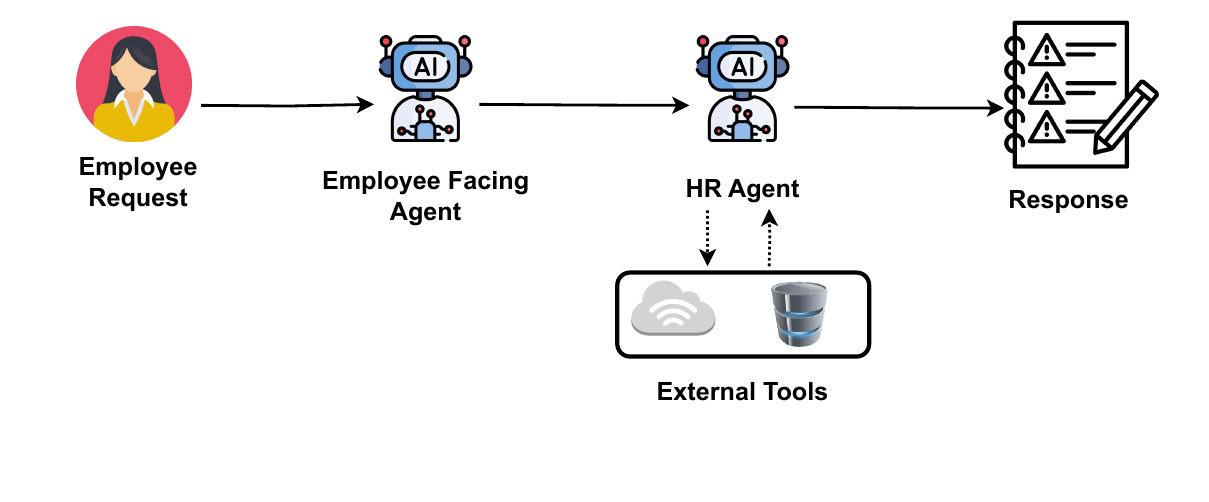}
    \caption{Overview of the HR assistant application.}
    \label{fig:hr}
\end{figure}

We evaluated this application under a prompt-injection threat model and demonstrated a data leakage vulnerability stemming from weak access control enforcement. Specifically, the system fails to implement strong authorization checks and context isolation when the agent orchestrates downstream actions such as retrieval or tool invocation. This allows injected instructions to induce the agent to access data outside the requesting user's permissions, resulting in cross-principal information disclosure where one employee's leave balances and bonus records are revealed to another employee. This scenario is particularly relevant for enterprise deployments where multi-tenancy and role-based access control are critical security requirements.

\noindent\textbf{Automated Trip Planner.}
We developed the multi-agent application design to autonomously generate a complete travel itinerary from a user’s trip request. It employs a sequential architecture comprising three specialized agents, each coupled with external tools. Upon receiving the user query, the city selection agent analyzes the request and extracts the essential trip constraints and preferences such as destination, travel dates or duration, and user interests, providing the inputs for downstream planning.
The travel research agent then performs in-depth research on the chosen city, compiling a comprehensive dossier that includes accommodation and attraction details, dining suggestions, practical local information, and approximate cost estimates. In the final stage, the itinerary generation agent consolidates the collected materials and transforms them into a structured, detailed itinerary.

\begin{figure}[h]
    \includegraphics[width=0.99\linewidth]{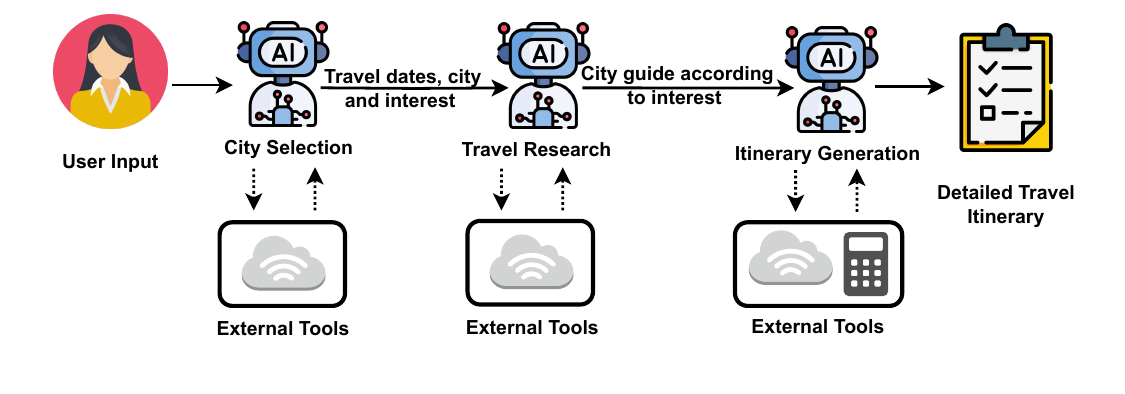}
    \caption{Overview of the automated trip planner application.}
    \label{fig:trip}
\end{figure}

A two-stage adversarial strategy for sensitive-context leakage in the automated trip planner is adapted from the ATAG study~\cite{gandhi2025atag}. In first stage, the attacker probes the system with multiple travel queries and studies structural regularities in the returned itineraries, including hyperlink placement, URL consistency, and the dependence of external links on query parameters. This behavior reveals a pipeline where an upstream research agent generates a structured city dossier that is consumed by downstream agents for itinerary construction. After identifying this boundary, the attacker compromises the inter-agent communication channel and selectively rewrites booking-link URLs with attacker-controlled domains while preserving all other dossier fields. Because the tampering preserves schema validity, it evades downstream structural checks. Lacking URL provenance validation for external links, the system propagates the modified URLs into the final itinerary. User interaction with these links leads to a spoofed booking interface enriched by leaked contextual data.

\noindent\textbf{Adapted ATAG Applications.}
We reused two multi-agent applications from the ATAG framework without modification to broaden the range of orchestration topologies, LLM backends, and attack surfaces in our evaluation. The automated email responder follows a hierarchical CrewAI architecture with GPT-4o, coordinating orchestrator, fetcher, categorizer, prioritizer, and drafter agents for end-to-end email triage and response. We examined prompt injections embedded in incoming emails that cause internal configuration leakage through inter-agent output propagation and data exfiltration via the reply channel. The MCP-based candidate hiring application uses CrewAI with Gemini-2.0-Flash and interfaces with external services through MCP servers for candidate evaluation and interview scheduling; we examined injections embedded in candidate-submitted documents that attempt to manipulate evaluation agents into disclosing other candidates' records. Full architectural details for both applications can be found in~\cite{gandhi2025atag}. 

\subsection{Ablation Study}
We isolate each module's contribution by removing it while retaining the other two (Table~\ref{tab:ablation}). The full pipeline is the reference: it attains $0\%$ ASR on four of five applications and $7.4\%$ on the Network Monitoring assistant (stress induced attacks only), while preserving BTSR.

\noindent\textbf{Analyzer.}
Without the analyzer, the modifier applies every transformation class uniformly, with no risk-prioritized targeting. Security degrades sharply because untargeted edits miss the application-specific leakage paths: ASR rises from $7.4\%$ to $31.2\%$ (Network Monitoring) and from $0\%$ to $36.8\%$ (Trip Planner). Utility also suffers from indiscriminate hardening of agents that handle no sensitive data; BTSR drops from $100\%$ to $88.0\%$ (Email Responder) and, most severely, from $76.9\%$ to $64.2\%$ on the HR assistant, whose sequential topology cascades an over-restrictive early-stage edit through all downstream stages. The analyzer therefore contributes to \emph{utility}: its audit report lets the modifier intervene precisely and leave benign flows intact.

\noindent\textbf{Modifier.}
Removing the modifier leaves the source unchanged, so BTSR stays at $100\%$ but ASR returns to pre-mitigation levels confirming that diagnosis without remediation yields no security benefit. The Network Monitoring and Trip Planner applications show the highest residual ASR ($70.8\%$ and $81.9\%$), reflecting the severity of their unpatched inter-agent and tool-mediated paths.

\noindent\textbf{Validation.}
Removing the validation module raises ASR by up to $20.4$\,pp (Network Monitoring)  as sophisticated adversarial attacks that evade the modifier's static edits go undetected without the validation module's runtime checks. BTSR also drops by $6.4$--$8.5$\,pp on four of the five applications, since modifier edits can silently break benign workflows that the validation module would otherwise detect and repair via iterative refinement. Validation thus serves a dual role: it catches residual  adversarial inputs that slip past earlier modules and preserves benign functionality, making it an active part of the hardening loop rather than a post-hoc check.

\begin{table*}[t]
\centering
\caption{Module ablation on the five CrewAI/LangGraph applications. Each column removes one module and keeps the other two.}
\label{tab:ablation}
\small
\begin{tabular}{llcccccccc}
\toprule
 & & \multicolumn{2}{c}{\textbf{Full pipeline}} & \multicolumn{2}{c}{\textbf{w/o Analyzer}} & \multicolumn{2}{c}{\textbf{w/o Modifier}} & \multicolumn{2}{c}{\textbf{w/o Validator}} \\
\cmidrule(lr){3-4}\cmidrule(lr){5-6}\cmidrule(lr){7-8}\cmidrule(lr){9-10}
\textbf{Framework} & \textbf{Application} & ASR$\downarrow$ & BTSR$\uparrow$ & ASR$\downarrow$ & BTSR$\uparrow$ & ASR$\downarrow$ & BTSR$\uparrow$ & ASR$\downarrow$ & BTSR$\uparrow$ \\
\midrule
\multirow{3}{*}{CrewAI}
  & Automated Email Responder & 0.0  & 100.0 & 22.4 & 88.0 & 50.3 & 100.0 & 3.4 & 93.5 \\
  & HR Assistant              & 0.0  & 76.9  & 29.7 & 64.2 & 79.7 & 100.0 & 2.8 & 68.4 \\
  & Candidate Hiring          & 0.0  & 97.1  & 17.6 & 81.3 & 38.1 & 100.0 & 12.1 & 90.7 \\
\midrule
\multirow{2}{*}{LangGraph}
  & Trip Planner Assistant    & 0.0  & 100.0 & 36.8 & 86.4 & 81.9 & 100.0 & 5.8 & 92.3 \\
  & Network Monitoring Assistant & 7.4 & 91.0 & 31.2 & 82.7 & 70.8 & 100.0 & 20.4 & 95.6 \\
\bottomrule
\end{tabular}
\end{table*} 
\section{Results}
 
We evaluate our hardening pipeline on the two dimensions: (1) the practical deployments must satisfy jointly first reducing leakage under adversarial injection, and (2) preserving benign utility. We measure both on five real-world applications (\ref{sec:results-casestudy}), on the AgentDojo benchmark (\ref{sec:results-agentdojo}), and against FIDES, a runtime information-flow-control (IFC) defense (\ref{sec:results-fides}).
 
\subsection{Leakage Reduction on Case-Study Applications}\label{sec:results-casestudy}
 
Table~\ref{tab:results} reports ASR before (B) and after (A) hardening for the five applications across four prompt-injection classes. Direct requests for protected data never succeeded $0\%$ ASR in every application, both before and after hardening. Leakage concentrates in attacks that re-prioritize or override how instructions are interpreted, not in explicit data requests. Stress-induced prompting is the strongest class in every application, peaking at $51.0\%$ ASR on CrewAI (HR Assistant) and $58.4\%$ on LangGraph (Network Monitoring). Instruction-override is moderate ($11.4$--$22.2\%$), and basic jailbreaks are weakest but non-trivial ($6.5$--$10.3\%$ on three applications).

After hardening, ASR falls to $0\%$ across all four classes for every CrewAI application and for the LangGraph Trip Planner a $100\%$ reduction in realized leakage. The sole residual is the LangGraph Network Monitoring assistant under stress, where stress-class ASR drops from $58.4\%$ to $7.4\%$ (an $87\%$ reduction \emph{on that class}); aggregated over all four classes, the application's leakage falls from $81.9\%$ to $7.4\%$, a $91\%$ reduction. Hardening thus eliminates the simpler vectors outright and sharply
attenuates the hardest one.
 
\begin{table*}[t]
\centering
\caption{ASR (\%) for four prompt-injection attack scenarios against five agentic AI applications built with CrewAI and LangGraph, measured before (B) and after (A) applying our mitigation pipeline.}
\label{tab:results}
\renewcommand{\arraystretch}{1.2}
\small
\resizebox{\linewidth}{!}{%
\begin{tabular}{l l c c c c c c c c c}
\hline
\multirow{2}{*}{\textbf{Framework}} &
\multirow{2}{*}{\textbf{Application}} &
\multicolumn{2}{c}{\textbf{Direct Request}} &
\multicolumn{2}{c}{\textbf{Basic Jailbreak}} &
\multicolumn{2}{c}{\textbf{Instruction Override}} &
\multicolumn{2}{c}{\textbf{Stress-Induced}} &
\multirow{2}{*}{\textbf{Reduction}} \\
\cmidrule(lr){3-4} \cmidrule(lr){5-6} \cmidrule(lr){7-8} \cmidrule(lr){9-10}
& & \textbf{B} & \textbf{A} & \textbf{B} & \textbf{A} & \textbf{B} & \textbf{A} & \textbf{B} & \textbf{A} & \\
\hline
\multirow{3}{*}{CrewAI}
  & Automated Email Responder  & 0.0 & 0.0 & 0.0 & 0.0 & 15.3 & 0.0 & 34.7 & 0.0 & 100\% \\
  & HR Assistant               & 0.0 & 0.0 & 6.5 & 0.0 & 22.2 & 0.0 & 51.0 & 0.0 & 100\% \\
  & Candidate Hiring           & 0.0 & 0.0 & 0.0 & 0.0 & 11.4 & 0.0 & 26.7 & 0.0 & 100\% \\
\hline
\multirow{2}{*}{LangGraph}
  & Network Monitoring Assistant & 0.0 & 0.0 & 10.3 & 0.0 & 13.2 & 0.0 & 58.4 & 7.4 & 91\% \\
  & Trip Planner Assistant     & 0.0 & 0.0 & 7.2 & 0.0 & 18.9 & 0.0 & 44.7 & 0.0 & 100\% \\
\hline
\end{tabular}%
}
\end{table*}

\subsection{Generalization to AgentDojo} \label{sec:results-agentdojo}
 
To test generalization beyond our own applications, we ran the full pipeline on all four AgentDojo suites under the \texttt{important\_instructions} attack and compared against AgentDojo's built-in defense, scoring with the benchmark's official, state-based utility and security checkers (no LLM judge). On the no-attack baseline both defenses reach $100\%$ utility and $100\%$ security, confirming that neither regresses benign behavior.
 
Under attack (Table~\ref{tab:agentdojo}), our approach yields a large utility advantage and a net security gain. Overall utility rises from $37.0\%$ (default) to $72.2\%$ (ours), $+35.2$\,pp, with improvement in every suite. The single regression is Slack security ($100.0\%\!\rightarrow\!80.0\%$,
$-20.0$\,pp). By keeping the agent operational rather than refusing on any suspicious content, we admit a few injections that the default blocks via
blanket refusal. 

\begin{table}[t]
\centering
\caption{AgentDojo under the \texttt{important\_instructions} attack:
AgentDojo's default defense vs.\ ours. U: utility (task success); S:
security (attack resistance). $\Delta$: ours $-$ default (pp).}
\label{tab:agentdojo}
\small
\begin{tabular}{lcccccc}
\toprule
\textbf{Domain} & \multicolumn{2}{c}{\textbf{Default}} & \multicolumn{2}{c}{\textbf{Ours}} & \multicolumn{2}{c}{$\boldsymbol{\Delta}$} \\
\cmidrule(lr){2-3}\cmidrule(lr){4-5}\cmidrule(lr){6-7}
 & U$\uparrow$ & S$\uparrow$ & U$\uparrow$ & S$\uparrow$ & $\Delta$U & $\Delta$S \\
\midrule
Workspace & 16.7 & 58.3  & 66.7 & 75.0 & $+50.0$ & $+16.7$ \\
Travel    & 28.6 & 28.6  & 71.4 & 64.3 & $+42.8$ & $+35.7$ \\
Slack     & 50.0 & 100.0 & 90.0 & 80.0 & $+40.0$ & $-20.0$ \\
Banking   & 50.0 & 77.8  & 66.7 & 77.8 & $+16.7$ & $\phantom{+}0.0$ \\
\midrule
\textbf{Overall} & 37.0 & 66.7 & 72.2 & 74.1 & $+35.2$ & $+7.4$ \\
\bottomrule
\end{tabular}
\end{table}

\subsection{Comparison with a Runtime IFC Defense: FIDES} \label{sec:results-fides}
 
Security alone does not capture defense quality: a mechanism can block leakage yet still break benign workflows, raise hallucination, or push the agent into unsafe fallback behavior. We therefore compare against FIDES~\cite{costa2025securingaiagentsinformationflow}, a recent planner-level IFC defense for prompt injection in tool-using agents. FIDES uses fundamentally different strategy of information-flow control and policy enforcement at the planner making it a strong reference point for measuring the \emph{practical cost} a defense imposes, not just its attack resistance. We stress that this is an assessment of trade-offs across designs, not a challenge to FIDES's contribution: FIDES shows that IFC provides strong formal protection. Our proposed pipeline is complementary in deployment; robustness must be evaluated jointly with functional preservation.
 
To capture failure modes that aggregate completion hides, Table~\ref{tab:fides} reports benign task success before and after mitigation (BTSR-B, BTSR-A) and the post-mitigation hallucination rate (unsupported outputs). A defense that is highly secure but overly restrictive merely relocates failure from unsafe tool use to lost utility and hallucinated fallbacks, a cost that is unacceptable in production where correctness and grounding are critical. 

\begin{table}[t]
\scriptsize
\centering
\caption{Comparative utility and failure analysis of modified FIDES and our mitigation pipeline.}
\label{tab:fides}
\resizebox{\columnwidth}{!}{%
\begin{tabular}{llccccc}
\toprule
\multirow{2}{*}{\textbf{Framework}} &
\multirow{2}{*}{\textbf{Application}} &
\multirow{2}{*}{\textbf{BTSR-B}} &
\multicolumn{2}{c}{\textbf{BTSR-A}} &
\multicolumn{2}{c}{\textbf{Hallucination-A}} \\
\cmidrule(lr){4-5} \cmidrule(lr){6-7}
& & & \textbf{FIDES} & \textbf{Ours} & \textbf{FIDES} & \textbf{Ours} \\
\midrule

\multirow{3}{*}{CrewAI}
  & Email Responder & 100 & 0    & 100  & 100.0 & 0    \\
  & HR Assistant    & 100 & 38.5 & 76.9 & 76.9  & 23.1 \\
  & Hiring          & 100 & 0    & 97.1 & 97.1  & 4.8  \\
\midrule
\multirow{2}{*}{LangGraph}
  & Network Monitor & 100 & 90.0 & 100 & 91.0  & 0   \\
  & Trip Planner    & 100 & 80.0 & 100 & 100.0 & 7.2 \\
\bottomrule
\end{tabular}%
}
\end{table}

Across both evaluation tracks the evidence points to the same conclusion, defenses for data leakage in LLM agents must be judged on two dimensions, simultaneously resistance to adversarial influence \emph{and} preservation of intended functionality. Our results suggest the second axis is decisive in applied systems, where overly conservative trust restrictions silently break benign workflows. 
\section{Conclusion}

In the proposed pipeline we have argued that data leakage in agentic applications is largely a build-time problem: in tool-using, multi-step systems, sensitive content propagates through prompt templates, memory, inter-agent messages, and tool arguments, producing leakage paths that prompt-level defenses cannot reliably close. Rather than policing leakage at runtime, we presented a pre-deployment pipeline that scans application source, hardens the highest-risk source-to-sink paths, and validates the result against adaptive adversarial and benign inputs.
Across five real-world CrewAI and LangGraph applications, hardening eliminated leakage under basic-jailbreak and instruction-override attacks (100\% reduction) and reduced it by 91\% under stress-induced setting, leaving 7.4\% ASR in the network monitoring assistant. On AgentDojo, the approach generalized to single-agent tool-calling and, critically, preserved utility while doing so it improved task completion under attack by 35.2 points over the default defense at a net +7.4 point security gain. This trade-off is our central finding: a defense that blocks adversarial actions by conservatively refusing benign ones merely relocates failure from unsafe tool use to lost utility and hallucinated fallbacks, a cost that is unacceptable in production.
A few limitations still remain: stress-framed prompts are reduced but not eliminated, schema-preserving tampering of inter-agent artifacts evades our structural checks, and multi-user deployments require stronger authorization and context isolation than we currently enforce. We therefore position build-time hardening not as a replacement for runtime defenses but as a complementary layer. Future work will pursue provenance for inter-agent communication, tighter authorization and memory scoping, and integration with runtime information-flow control to achieve defense-in-depth across the runtime boundary.

\bibliographystyle{IEEEtran}
\bibliography{references}

\end{document}